# Two-Step Adsorption of $PtCl_6^{2-}$ Complexes at a Charged Langmuir Monolayer: Role of Hydration and Ion Correlations

Ahmet Uysal,[1,][*] William Rock,[1] Baofu Qiao,[1] Wei Bu,[2] and Binhua Lin[2]

[1]Chemical Sciences and Engineering Division, Argonne National Laboratory, Argonne IL 60439

[2]Center for Advanced Radiation Sources, The University of Chicago, Chicago IL 60637

AUTHOR INFORMATION

**Corresponding Author**

*E-mail: ahmet@anl.gov.

Anion exchange at positively charged interfaces plays an important role in a variety of physical and chemical processes. However, the molecular scale details of these processes, especially with heavy and large anionic complexes, are not well-understood. We studied the adsorption of $PtCl_6^{2-}$ anionic complexes to floating DPTAP monolayers in the presence of excess $Cl^-$ as a function of the bulk chlorometalate concentration. In situ x-ray scattering and fluorescence measurements, which are element and depth sensitive, show that the chlorometalate ions only adsorb in the diffuse layer at lower concentrations, while they adsorb predominantly in the Stern layer at higher concentrations. The response of DPTAP molecules to the adsorbed ions is determined independently by grazing incidence x-ray diffraction, and supports this picture. Molecular dynamics simulations further elucidate the nanoscale structure of the interfacial complexes. The results suggest that ion hydration and ion-ion correlations play a key role in the competitive adsorption process.

**TOC GRAPHICS**

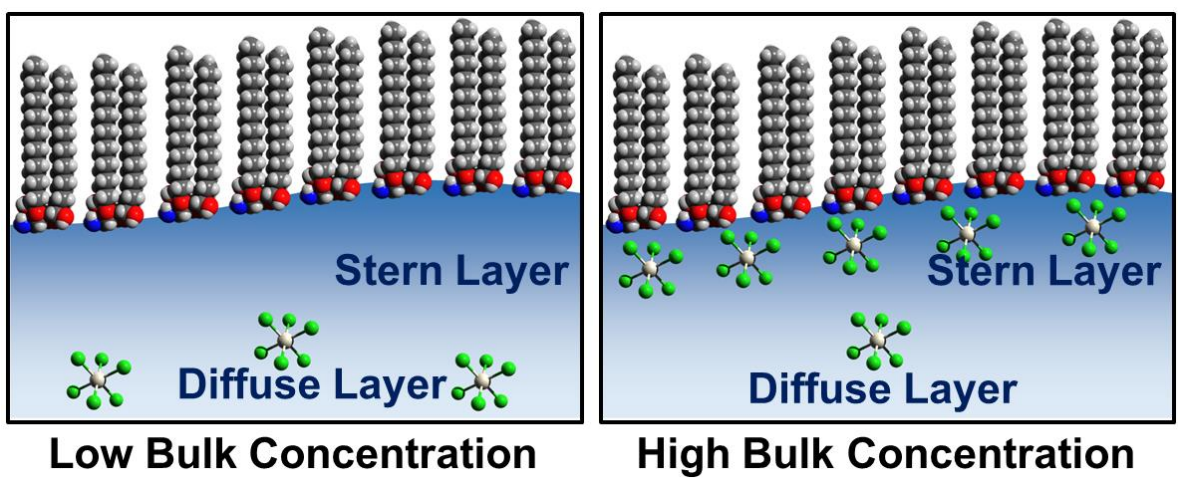

Amphiphile-ion interactions at aqueous interfaces play an important role in many biological, chemical, environmental and industrial processes.[1-3] For instance, nearly a third of the world's copper production relies on solvent extraction (SX),[3] a chemical separations technique in which targeted metal ions are transferred from an aqueous phase, through an interface, and into an immiscible organic phase with the help of amphiphilic extractants. SX is also the main technique used in the refining and reprocessing of precious and rare earth metals[4-5] and nuclear waste.[6] Although SX has been developed since the mid twentieth century,[7] very little is known about the molecular-scale interactions during the transfer of the ions through an aqueous/oil interface; these important interactions are expected to govern SX kinetics.[8-11] A detailed, molecular-level understanding of the interactions and structures that are present during metal-ion transfer is required to predict and develop SX and other solution processes to meet the high demands of modern technologies.[4]

A key gap in the understanding of SX is a clear picture of the structure of interfacial ions in the aqueous phase at the oil/water interface.[10-11] These ions can be modeled as an electric double layer (EDL) at a charged interface, which has been the subject of theoretical and experimental studies for more than a century.[12-14] The early version of Gouy-Chapman (GC) theory, which assumes dilute solutions with point charges in a continuous dielectric environment, can explain many experimental observations for monovalent ions.[12] Later improvements to GC theory include correction factors for multivalent ions, the finite size of the ions, short range interactions, ion-ion interactions, and non-uniform dielectric constants.[12] However, our theoretical understanding is still being challenged by new experimental results, especially at high ion concentrations, and when specific ion and surface effects are important.[15-16] Also, large anionic heavy metal complexes are not addressed in most interfacial studies.[17] Within this context, interface specific experimental studies of heavy metal ions, such as platinum group metals (PGMs) are required to explore the limits of our theories, and to improve the technologies that rely on them.

The behavior of anionic complexes of PGMs can be a good model for radioactive waste in high halide environments, such as $PuCl_6^{2-}$.[18-19] It is suggested that the hydrated radius of anions plays an important role in their surface interactions: the larger the anion, the more it is attracted to the surface.[17] Actinide and PGM anionic complexes generally have thermochemical radii greater than 3 Å, while most surface studies have been done with lighter anions with thermochemical radii less than 2.5 Å.[20] Therefore, experiments are necessary in this traditionally unexplored regime.

Advancements in interfacial probes, such as interfacial x-ray and neutron scattering,[9-12, 21-22] nonlinear vibrational spectroscopy,[8, 23-25] and surface force measurements,[15] dramatically increased our molecular-scale understanding of ions at charged interfaces in recent years. Extractant-ion complexes at the oil/water interface have been investigated with *in situ* scattering and spectroscopy experiments.[9-11] Also, experiments designed to model certain aspects of ion speciation[26-28] and extractant-ion interactions at the air/water[29-30] and solid/water[22] interfaces have been reported.

Recently, it has been shown that anionic complexes of platinum group metals, $PtCl_6^{2-}$ and $PdCl_4^{2-}$, can adsorb on amine functionalized surfaces even if their concentrations are four orders-of-magnitude smaller than the background $Cl^-$ concentrations.[22] These experiments addressed an apparent dichotomy between mean-field theories of the competitive adsorption of ions with different valencies, and the SX of chlorometalates from highly concentrated chloride solutions. It was suggested that the pure Coulombic nature of the mean-field theories fails to describe the competition in aqueous environments since ion hydration and other short-range interactions actually play very significant roles. It was also suggested, by indirect observations, that at a relative concentration ($[\text{metalate}^{2-}]/[\text{Cl}^-]$) of $10^{-4}$, the metalate ions adsorb in the diffuse layer, but not in the Stern layer. The metalate ions were adsorbed in the Stern layer at higher relative concentrations.

In this Letter, we study the adsorption of $\text{PtCl}_6^{2-}$ anions at 1,2-dipalmitoyl-3-trimethylammonium-propane (DPTAP) monolayers at the air/water interface (Figure 1). Having a quaternary amine head-group, DPTAP is a good model for most common extractants used in PGM extraction.[31-32] Also, its interactions with lighter anions are well-documented.[33-34] The air/water interface is a good model system to study the aqueous side of the oil/water interface, since air has a similar dielectric constant ($\varepsilon = 1$) to oil ($\varepsilon \sim 2$), and allows us to tune the relevant parameters more easily.[26-30] Our aqueous subphase contains 0.5 M LiCl to simulate the process conditions, and the $[\text{PtCl}_6^{2-}]/[\text{Cl}^-]$ ranges from $10^{-5}$ to $10^{-1}$. Anomalous x-ray reflectivity (a-XR) and x-ray fluorescence near total reflection (XFNTR) measurements directly determine the amount of $\text{PtCl}_6^{2-}$ complexes adsorbed in the diffuse and Stern layers, respectively. These methods provide a detailed, direct measurement of the two-step adsorption process, which was only indirectly observed at the solid/liquid interface.[22] The structural changes in the DPTAP film in response to the adsorbed ions were determined by grazing incidence x-ray diffraction (GID) as a function of the bulk metalate concentration, and they also correlate with the two-step adsorption picture determined by a-XR and XFNTR. Molecular dynamics simulations provide further information on the nanoscale DPTAP-metalate structures at the air/water interface.

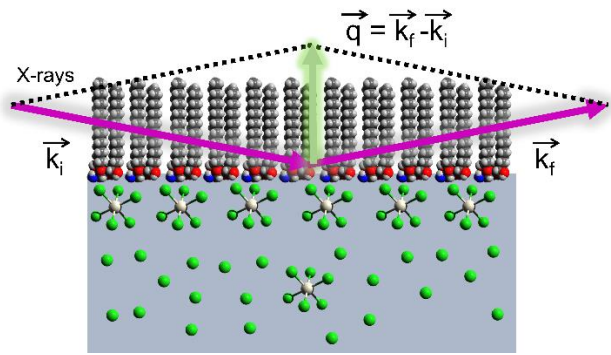

**Figure 1.** A schematic of the specular x-ray reflectivity experiments at the air/liquid interface. DPTAP molecules and adsorbed $PtCl_6^{2-}$ ions create two interfacial layers with different electron densities. The thickness and the density of these layers can be determined from the x-ray reflectivity data.

We first determine the metalate concentration in the Stern layer with a-XR measurements. Figure 2a shows the Fresnel normalized reflectivity data (symbols) for concentration dependent a-XR measurements. At each concentration, XR measurements were done on the $L_3$ absorption edge (11.564 keV) of Pt (blue), and 250 eV below the edge (red). Because of x-ray absorption effects, the effective number of Pt electrons that scatter x-rays decreases by 16 on the edge (blue), compared to the measurement done below the edge (red) (Figure S1, Supporting Information). The rest of the system should not be affected by the change in the x-ray energy.[12] Therefore, the differences observed between the red and the blue XR data, at a specific concentration, are only due to the presence of the $PtCl_6^{2-}$ ions. The oscillations in the XR data (Kiessig fringes) are due to the electron density gradient normal to the interface, caused by the DPTAP monolayer and the adsorbed ions in the Stern layer forming two distinct layers between air and the subphase solution. The increase in the amplitude of these oscillations and the shifting of the minima towards smaller q values clearly show that both the electron density contrast between the layers and the overall thickness of the interfacial layers increase with increasing chlorometalate concentration. A shift of the off-edge data (red) relative to the on-edge (blue) data indicates that some of the interfacial electron density is due to the adsorption of Pt ions – qualitative evidence that Pt ions contribute to the increase in interfacial electron density. These observations can be quantified by a model-dependent fit of the XR data.

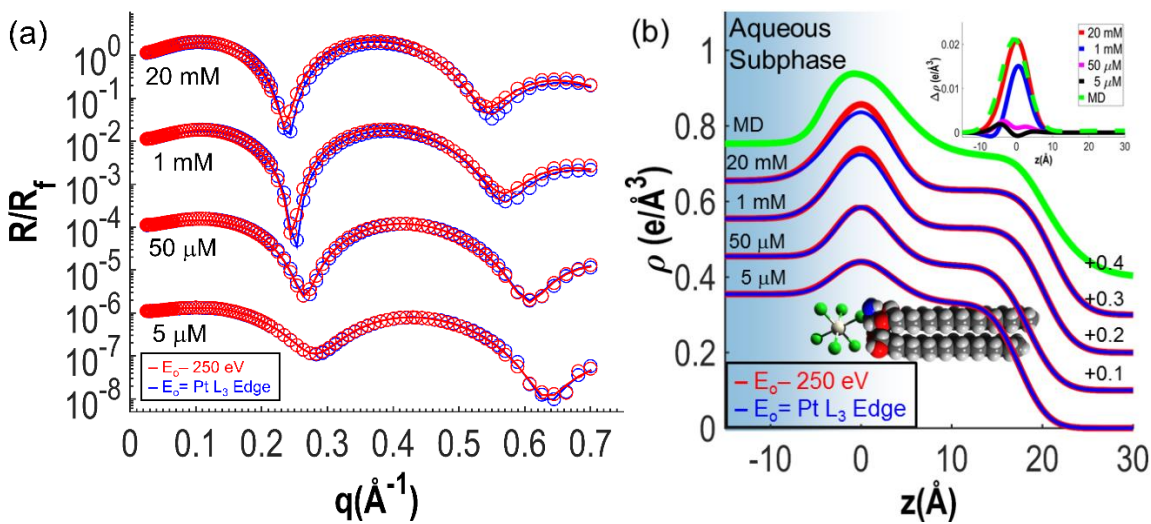

**Figure 2. (a)** Anomalous XR data (symbols) and fits (solid lines) of DPTAP monolayers on a subphase with 0.5 M LiCl and varying $PtCl_6^{2-}$ concentrations. The plot contains XR measurements at the platinum $L_3$ absorption edge (blue, 11.564 keV) and 250 eV below the edge (red). Data for each concentration is shifted by two decades for clarity. **(b)** Electron density profiles (EDPs) derived from the fits to the XR data in (a) (blue and red) and from the MD simulations (green). EDPs at each concentration are shifted by 0.1 e/Å$^3$ for clarity. The cartoon depicts the approximate positions of DPTAP molecules and $PtCl_6^{2-}$ ions at the air/water interface corresponding to the EDPs. The inset shows the difference between the red (off edge) and the blue (on edge) EDPs for each concentration. The differential EDPs in the inset are compared to an appropriately scaled $PtCl_6^{2-}$ distribution from MD simulations (dashed green curve).

The fits model the interface as two laterally homogenous layers between the air and the subphase (Figure 2b, cartoon). One box corresponds to the tail region of the DPTAP, while the other box is assigned to the headgroup and the ions in the Stern layer. Nonlinear fitting of the XR data with the calculated XR curves from this model (Figure 2a, lines) provides the thickness, electron density, and roughness parameters (SI-Table S1, Supporting Information) for these layers,[35-36] which are used to plot the EDPs in Figure 2b. The difference between the red and the blue curves for each concentration is solely due to the Pt in the $PtCl_6^{2-}$ complex. The inset shows the differential EDPs, calculated by subtracting on-edge EDPs (blue) from off-edge (red) EDPs. As expected, the $PtCl_6^{2-}$ coverage increases with increasing bulk concentration. The thickness of the tail groups also increases with increasing metalate adsorption, meaning that the tilt angle (measured from the surface normal) of the DPTAP tails decreases with increasing metalate concentration.

The calculated EDPs from MD simulations are also shown in Figure 2b (green). To determine the maximum coverage structures, the $[PtCl_6^{2-}]$ is increased in the simulations until the surface is saturated, and full simulations are run at that concentration. As shown in inset of Figure 2b, the $PtCl_6^{2-}$ EDP obtained from MD simulations closely matches the 20 mM differential EDP, suggesting that the Stern layer has reached its maximum coverage at this concentration. Considering that the thickness of the Stern layer is approximately 5 Å (Figure S3, Supporting Information), the coverage calculated from the differential EDP corresponds to ~3.6 M $[PtCl_6^{2-}]$ in that region.

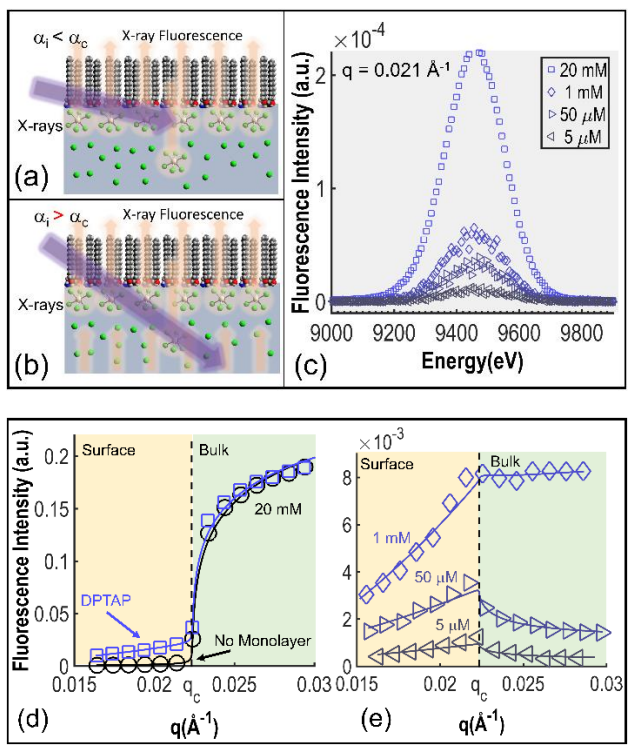

**Figure 3.** A depiction of XFNTR measurements below (a) and above (b) the critical angle. While only surface adsorbed ions (both diffuse and Stern layers) fluoresce below the critical angle, the ions in the bulk are also excited above the critical angle. (c) Concentration dependent fluorescence signal at fixed q = 0.021 Å$^{-1}$ below the critical angle. (d, e) Concentration dependent XFNTR data (symbols) calculated by integrating the area under the curves in (c) and their counterparts at various q values. Solid lines show the fits to the data as described in the text. 20 mM sample in (d) is measured with (squares) and without (circles) DPTAP monolayer. All measurements in (e) are done with DPTAP.

We can also quantify the total PtCl$_6^{2-}$ concentration, in both the Stern and diffuse layers, with XFNTR. Figure 3a depicts incoming x-rays with an incidence angle less than the critical angle. These x-rays do not penetrate into the bulk due to total external reflection,[37] and excite only the ions within 7-10 nm of the surface. If the incidence angle is above the critical angle, then the x-rays penetrate several microns into the bulk solution (Figure 3b). The fluorescence signal is measured with an energy dispersive detector 1 cm above the surface. Figure 3c shows the fluorescence signal measured at the Pt L$_\alpha$ emission line, at various concentrations, at an incidence angle that corresponds to q = 0.021 Å$^{-1}$. By varying the incidence angle slightly below and above the critical angle, we can determine the total fluorescence signal as a function of q (Figure 3d and e). Figure 3d shows the XFNTR data for the 20 mM sample with (circles) and without (squares) the DPTAP monolayer. The data below the critical angle (left side) is only sensitive to the surface adsorbed ions.

As expected, there is no surface signal without the monolayer (Figure 3d, left side). This data set is used to calibrate the fluorescence signal from the known bulk concentration. In the presence of the DPTAP monolayer, all concentrations display a surface-sensitive fluorescence signal below the critical angle (Figure 3 d and e). This signal linearly increases with q because the transmission factor linearly increases below the critical angle.[37-38] By fitting these data sets we can calculate the total coverage of PtCl$_6^{2-}$ ions at the interface (Figure 4).

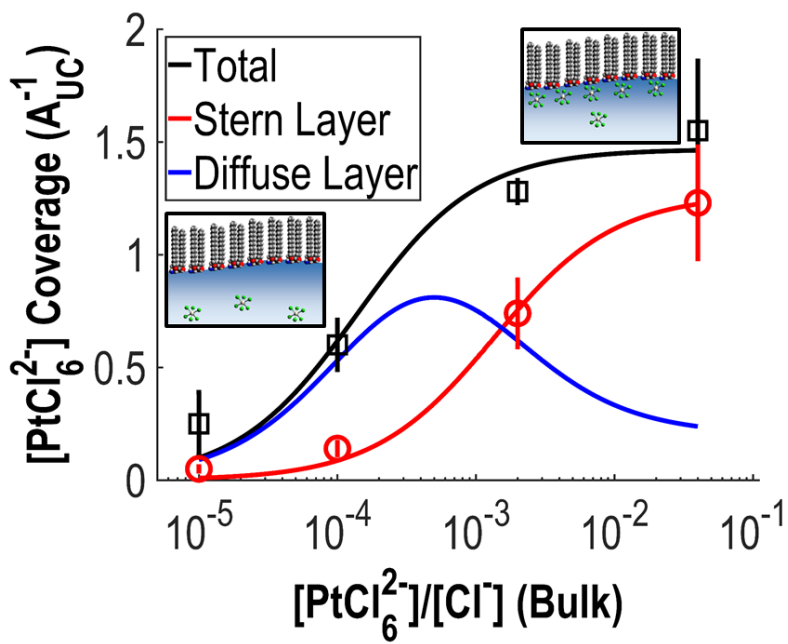

**Figure 4.** Coverage of $PtCl_6^{2-}$ ions as a function of the bulk concentration. The total coverage is calculated from XFNTR measurements (black squares). The contribution of the $PtCl_6^{2-}$ ions in the Stern layer is calculated from a-XR measurements (red circles). The solid black and red lines are Langmuir adsorption fits to the corresponding data sets. The diffuse layer contribution (blue line) is calculated as the difference between the fits to the total and the Stern layer coverage. The inset cartoons visualize the adsorption behavior in the Stern and diffuse layers at low and high bulk concentrations.

Figure 4 shows the concentration dependent interfacial coverage ($\theta$) of $PtCl_6^{2-}$ ions calculated from a-XR and XFNTR. Full coverage ($\theta = 1\ A_{UC}^{-1}$) is defined as the coverage when there is 1 $PtCl_6^{2-}$ anion per 2 DPTAP molecules at the interface (1 $A_{UC}$ is 100.3 Å$^2$, which has 2 DPTAP molecules, at the lowest concentration). The difference between the coverages calculated from the two methods is due to the fact that a-XR is only sensitive to the ions adsorbed in the Stern layer ($\theta_{XR} = \theta_{Stern}$), while XFNTR probes all the interfacial ions, both in the Stern layer and in the diffuse layer ($\theta_{XFNTR} = \theta_{Stern} + \theta_{Diffuse}$).[11] Therefore, the combination of these techniques provides a unique opportunity to identify what portion of the adsorbed ions that are directly interacting with the DPTAP headgroups. Coverages greater than 1 mean that chlorometalate complexes overcharge the surface at higher bulk concentrations. The MD simulations (Figure S3, Supporting Information) confirm the overcharging, and show that charge neutrality is satisfied by the positively charged counterions forming another layer under the $PtCl_6^{2-}$ layer.

The response of the DPTAP monolayer to the adsorbed ions can be quantified in detail by GID measurements. Figure 5a shows a typical GID pattern with one in-plane and one out-of-plane peak. The GID patterns for all concentrations are qualitatively similar; only the peak positions change with increasing concentration. The inset in figure 5a shows the concentration-dependent peak shift of the out-of-plane peak. These peaks are the signature of an NN (nearest neighbor) tilted monolayer; the molecular area and the tilt angle of the tails can be calculated from the peak positions (Figure 5b).[39] The tilt angle, measured from the surface normal, decreases from 38° to 33° with increasing concentration. This agrees with the a-XR measurements, which show that the thickness of the film in the z direction increases with increasing concentration (Figure 2b). The molecular area per DPTAP also decreases with increasing metalate concentration. The decrease in the molecular area occurs because the electrostatic repulsion between the DPTAP headgroups decreases when anions adsorb in the Stern layer (recall that all experiments were done at constant surface pressure).[17] The molecular area of DPTAP molecules stays constant at lower concentrations, and decreases when the relative metalate concentration is above $10^{-4}$. This trend correlates well with the measured ion coverage in the Stern layer, as it stays constant until the relative concentration is $10^{-4}$, then increases at higher bulk concentrations. (Figure 4).

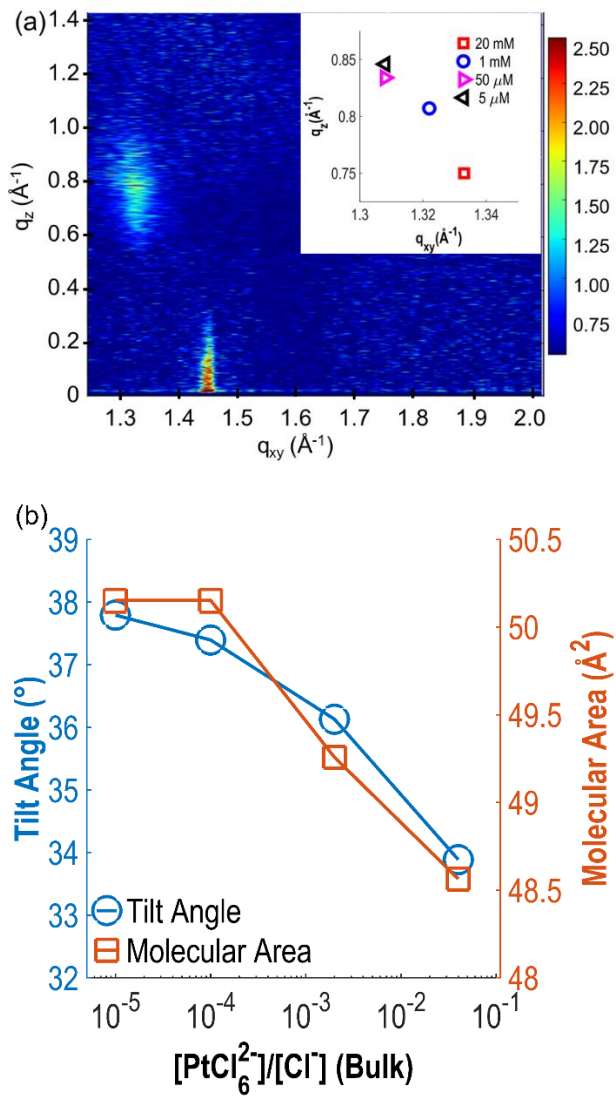

**Figure 5.** (a) GID data from DPTAP on a 1mM $PtCl_6^{2-}$ solution. One in-plane and one doubly degenerate out-of-plane peak is the signature of nearest neighbor (NN) tilt. The position of the out-of-plane peak changes with the concentration as shown in the inset. (b) Tilt angle of the DPTAP molecules from the surface normal (blue circles, left axis) and their molecular area (orange squares, right axis) as a function of the metalate concentration.

These results reveal interesting adsorption behavior for $PtCl_6^{2-}$ complexes in the presence of 0.5 M LiCl. If $[PtCl_6^{2-}]/[Cl^-]$ is less than $10^{-4}$, a-XR measures negligible $PtCl_6^{2-}$ adsorption in the Stern layer (Figure 2). Also, GID shows that the molecular area of the DPTAP monolayer does not change in this regime (Figure 5), supporting the idea that there is minimal direct interaction between the DPTAP headgroups and $PtCl_6^{2-}$ ions. However, XFNTR measures > 0.1 M [$PtCl_6^{2-}$] within 7-10 nm of the surface, which is ~1000 times higher than the bulk concentration. Recall that XFNTR detects all the interfacial $PtCl_6^{2-}$, and that a-XR measures $PtCl_6^{2-}$ only in the Stern layer,[11] therefore: $\theta_{Diffuse} / (\theta_{Stern} + \theta_{Diffuse}) = (\theta_{XFNTR}-\theta_{XR}) / \theta_{XFNTR}$. At a relative concentration of $10^{-4}$, 80% of all adsorbed chlorometalate ions are in the diffuse layer. $\theta_{Diffuse}$ reaches a maximum around a relative concentration of $3\times10^{-4}$, then starts dropping. At a relative concentration of $4\times10^{-2}$, only 20% of the interfacial metalates are in the diffuse layer. These observations suggest that the interfacial free energy landscape is significantly affected by the changes in the bulk chlorometalate concentration.

Hydration and interfacial water restructuring can create complicated free energy profiles at charged interfaces.[21] However, it is unexpected that the bulk ion concentration can alter that profile. A previous study measured the simultaneous adsorption of divalent $Sr^{2+}$ ions in the Stern and diffuse layers. In that work, the $Sr^{2+}$ ions in Stern layer did not compensate the surface charge completely, and Gouy-Chapman theory accurately modeled the relative concentrations of ions in Stern and diffuse layers.[11] In the present experiments the large $Cl^-$ excess creates a delicate balance in adsorption energetics, which favors $PtCl_6^{2-}$ adsorption in either the Stern layer or diffuse layer depending on the bulk concentration. The most important energetic consideration is likely the competition between the electrostatics and hydration. The Gibbs free energy of hydration ($\Delta G°_{hyd}$) of $PtCl_6^{2-}$ is -685 kJ/mol, which is ~50 % stronger than $F^-$ (-465 kJ/mol).[40] It is known that, due to its strong hydration, $F^-$ does not adsorb in Stern layer.[24] Therefore, it is reasonable to expect $PtCl_6^{2-}$ to only adsorb in diffuse layer. However, being a divalent anion, $PtCl_6^{2-}$ has a stronger electrostatic attraction to the surface than $F^-$.[41]

While hydration and electrostatics are the main drivers, there are at least two more important factors shaping the free-energy landscape: excess $Cl^-$ and ion-ion correlations. First, consider the low [$PtCl_6^{2-}$] regime; in this regime, there is little measured $PtCl_6^{2-}$ Stern layer adsorption, and $Cl^-$ is not expected to adsorb in the Stern layer[33], so both $PtCl_6^{2-}$ and $Cl^-$ likely retain their hydration spheres. The $\Delta G°_{hyd}$ of $Cl^-$ (-350 kJ/mol) is much weaker than $PtCl_6^{2-}$, so if hydration effects dominate the adsorption behavior in this regime, $Cl^-$ will adsorb closer to the surface, and screen $PtCl_6^{2-}$ from the surface charge. Diffuse layer adsorption is more favorable when the $PtCl_6^{2-}$ coverage ($\theta_{XFNTR}$) is < 1 (Figure 4), and interfacial $Cl^-$ ions contribute to the surface charge compensation. The [$PtCl_6^{2-}$] in the diffuse layer reaches its maximum when $\theta_{XFNTR}$ is ~ 1, which means that $PtCl_6^{2-}$ completely compensates the surface charge, and $Cl^-$ is excluded from the interface. At this point, electrostatic interactions overcome hydration and pull the $PtCl_6^{2-}$ ions into Stern layer. Therefore, Stern layer adsorption dominates when $\theta_{XFNTR} > 1$.

Ion-ion correlations are also expected to play an important role in $PtCl_6^{2-}$ adsorption. This contribution is quantified as a coupling strength $\Gamma = Z^2 l_B/d$, where Z is the ionic charge, d is the distance between the ions and $l_B = e^2/(\varepsilon k_B T)$ is Bjerrum length, the intermolecular distance at which electrostatic energy between two elemental charges (e) in a dielectric medium with permittivity ε is equal to their thermal energy $k_B T$.[14, 41-42] At a relative concentration of $10^{-4}$, $PtCl_6^{2-}$ ions are mainly adsorbed in the diffuse layer (XFNTR measures up to 70-100 Å depth, but the diffuse layer may be shorter), which corresponds to an interfacial concentration of ~0.1 M and a coupling strength $\Gamma$ ~ 1.1. This coupling strength is not high enough to cause significant overcharging.[42] At a relative concentration of $4\times10^{-2}$, assuming all the $PtCl_6^{2-}$ detected by XFNTR are distributed homogenously within the detected region (which is an underestimate), the coupling strength is $\Gamma$ ~ 2. This increase in the coupling strength, and the loss of shielding from $Cl^-$ described above, allows $PtCl_6^{2-}$ ions to overcome the energy barrier to lose half of their hydration shell and adsorb in the Stern layer at higher bulk concentrations. Also, it is important to note that at higher concentrations $PtCl_6^{2-}$ ions mostly adsorb in the Stern layer (~ 5 Å depth), which corresponds to $\Gamma$ ~ 3.6. This relatively larger coupling strength explains the overcharging, i.e. presence of ~ 50 % more $PtCl_6^{2-}$ than is needed to compensate the surface charge (0.32 $C/m^2$) due to the DPTAP head groups. Finally, in coupling strength calculations, it was assumed that the relative permittivity of water is constant at $\varepsilon$ ~ 78 at all concentrations. However, it is well-known that the interfacial water structure may change due to the adsorbed ions, which may cause the relative permittivity to drop significantly.[41] If $\varepsilon$ is smaller than 78, the ion-ion correlations might be even stronger than what was calculated.

The combination of the a-XR, XFNTR, and GID experiments integrated with the MD simulations provides a detailed picture of $PtCl_6^{2-}$ adsorption at DPTAP monolayers in the presence of excess $Cl^-$. At low concentrations, chlorometalate anions adsorb predominantly in the diffuse layer due to hydration effects. As $[PtCl_6^{2-}]/[Cl^-]$ increases, Stern layer adsorption becomes dominant, due to a combination of effects including interfacial $Cl^-$ deficiency, ion-ion correlations, and possibly a change in the interfacial water structure. Our results also show that in process conditions, when various ions with different valencies, sizes and hydration strengths interact, a detailed investigation with multiple probes becomes necessary to elucidate the details of the process. Although some aspects of these interactions were inferred from XR measurements previously,[22] the detailed and quantitative picture presented here was only possible with the combination of element specific scattering and x-ray fluorescence techniques combined with MD simulations.

These results provide important insights about the SX of chlorometalates from chloride solutions. Interfacial anion exchange is usually the rate limiting step in those processes.[32] Our results show that in highly concentrated solutions, ion-ion correlations and the competition between the co-ions are coupled with hydration effects, and are all important in determining adsorption behavior (the complexation of ions with headgroups can also lead to ion specific effects[17, 43-44], which is not the case here). These effects and the restructuring of interfacial water are expected to play a crucial role in SX processes. It is well-known that slight changes in process conditions, such as pH or temperature, may have significant effects on the kinetics of SX processes.[8-11] This study is a systematic investigation of the role of multiple factors in such challenging situations. A fundamental understanding of these factors will help to develop better separations techniques necessary for future energy technologies.

**Experimental and Computational Methods**

Anhydrous lithium chloride (LiCl, 99%), chloroplatinic acid solution ($H_2PtCl_6$, 8 wt. % in $H_2O$), and HPLC grade chloroform ($CHCl_3$, $\geq$ 99.9%) were purchased from Sigma-Aldrich. 1 N hydrochloric acid (HCl) was purchased from Fisher Scientific. 1,2-dioleoyl-3-trimethylammonium-propane (DPTAP) chloride salt was purchased in powder form from Avanti Polar Lipids and stored at -20° C. All chemicals were used as received. All subphase solutions contain 500 mM LiCl, and are adjusted to pH 2 using 1 N HCl.

The Langmuir monolayer is prepared using drop-wise addition of 0.25 mM DPTAP in $CHCl_3$ from a 100 μL Hamilton syringe. A Nima pressure sensor (from a model 601A Langmuir trough) using a chromatography paper Wilhelmy plate measures the surface pressure of the Langmuir monolayer. All experiments are performed at 8° C and at a constant surface pressure of 10 mN/m.

Synchrotron x-ray experiments were done at Sector 15-ID-C of the Advanced Photon Source at Argonne National Laboratory.[45] The x-ray energy was tuned around the Pt $L_3$ edge (11.565 ± 0.250 keV) depending on the experiment as described in the text. Two pairs of motorized slits set the incident beam size to 2 mm horizontally and 0.02 mm vertically. A Pilatus 100K area detector records the scattered x-ray signal, and a Vortex-60EX multicathode energy dispersive x-ray detector, placed perpendicularly 10 mm above the surface, records the x-ray fluorescence signal. The sample chamber was purged with helium to reduce the beam damage and the background scattering. The sample was shifted perpendicular to the beam, periodically, to avoid any beam damage due to long x-ray exposure.

In XR measurements, the specularly reflected x-ray intensity was recorded as a function of the vertical momentum transfer $|\vec{q}| = (4\pi/\lambda) \sin(2\theta/2)$, where $\lambda$ (1.07 Å at 11.564 keV and 1.10 Å at 11.314 keV) is the wavelength, and $\theta$ is the incidence angle (Figure 1). The electron densities of the films are modeled by two slabs, one for the tail region, and the other for the head group plus the adsorbed ions. The thickness, electron density and roughness of these layers are determined by least square fitting of the XR data to the calculated XR curves according to the Parratt formalism (Table S1, Supporting Information).[35-36, 38] The XR data at two different energies for a particular concentration were fitted simultaneously with the same parameters, except the headgroup density and the roughness between the headgroup and the subphase.

In the XFNTR measurements, the x-ray energy was fixed at 11.814 keV, which is above the $L_3$ absorption edge of Pt. The x-ray fluorescence intensity (Pt $L_\alpha$ emission at 9.442 keV) is recorded as a function of the vertical momentum transfer, $|\vec{q}|$, which is a function of the incidence angle (Figure 3a). The volume of the liquid illuminated by the x-rays is calculated from the beam dimensions as described in detail previously.[11, 37] The foot print of the beam on the liquid surface was always larger than the detector, making only the depth of the illuminated volume a function of the incidence angle.

In the GID measurements, the x-ray energy was fixed at 11.314 keV. The incidence x-ray angle was fixed to 0.019 Å$^{-1}$, and the detector was moved in the plane of water surface to record the diffraction patterns. Only a 3 pixel wide (~0.5 mm) stripe of the area detector was used in diffraction pattern reconstruction (Figure 5a) to obtain high $q_{xy}$ resolution; $q_z$ resolution was defined by the pixel size (172 μm). The peak positions are determined from linear plots obtained by vertical and horizontal integration of the diffraction patterns. The molecular areas and the tilt angles are calculated form these peak positions.[39]

Classical molecular dynamics (MD) simulations were performed using the GROMACS package (version 4.5.5).[46] The CHARMM 36 force field[47] was employed, which has been implemented under the GROMACS package [48]. The force field parameters of $PtCl_6^{2-}$ have been reported previously [49]. All other details of the setup, materials, XR fit parameters, and MD simulations are documented in the supporting information.

ASSOCIATED CONTENT

**Supporting Information**. XR fitting methods and fit parameters, MD simulation methodology and details.

AUTHOR INFORMATION

**Notes**

The authors declare no competing financial interests.

ACKNOWLEDGMENT

We thank Lynda Soderholm for her valuable comments on the manuscript. We also thank Sang Soo Lee for fruitful discussions. This work was supported by the U.S. Department of Energy, Office of Basic Energy Science, Division of Chemical Sciences, Geosciences, and Biosciences, Heavy Element Chemistry, under contract DE-AC02-06CH11357. The synchrotron x-ray experiments were done at ChemMatCARS, Sector 15-ID-C of the Advanced Photon Source at Argonne National Laboratory. Use of the Advanced Photon Source was supported by the U.S. Department of Energy, Office of Basic Energy Science, under contract DE-AC02-06CH11357. ChemMatCARS was supported by NSF/CHE-1346572. The MD simulations were done at the computing resources provided on Blues, a high-performance computing cluster operated by the Laboratory Computing Resource Center at Argonne National Laboratory.

**Supporting Information**

# Two-Step Adsorption of PtCl$_6^{2-}$ Complexes at Charged Langmuir Monolayers at the Air/Water Interface

Ahmet Uysal,[1, *] William Rock,[1] Baofu Qiao,[1] Wei Bu,[2] and Binhua Lin[2]

[1]Chemical Sciences and Engineering Division, Argonne National Laboratory, Argonne IL 60439
[2]Center for Advanced Radiation Sources, The University of Chicago, Chicago IL 60637

**a-XR Data Fitting**

The x-ray reflectivity for a known electron density profile (EDP) is calculated by

$$R(q) = R_F \left| \frac{1}{\Delta\rho} \int \frac{d\rho(z)}{dz} e^{-iz\sqrt{q(q^2-q_c^2)^{\frac{1}{2}}}} dz \right|^2$$

Here $R_F$ is the Fresnel reflectivity, i.e. the ideal reflection from an interface with zero roughness; $q_c$ is the critical angle and $\Delta\rho$ is the electron density change through the interface.[1-2] To determine an unknown EDP from a known R(q), we model the interface with constant density slabs with error function interfaces:

$$\rho(z) = \rho_0 + \sum_{i=0}^{n} \frac{\rho_{i+1} - \rho_i}{2} \left[ 1 + \text{erf}(\frac{z - z_i}{\sqrt{2}\sigma_i}) \right]$$

Here $\rho_i$ and $\sigma_i$ are the electron density and the roughness of i$^{th}$ slab, respectively. We assign one layer to the head group of DPTAP and the adsorbed ions, and another layer to the tail group. Our fits determine the length (L), electron density ($\rho$) and interfacial roughness ($\sigma$) of the each layer (Table S1). Data from on-edge ($E_o$=11.564 keV) and off-edge ($E_o$-250 eV=11.314 eV) measurements at the same bulk concentration are simultaneously fit with all parameters linked except the $\rho$ and $\sigma$ in the headgroup-ion region; all other parameters will not be affected by a change in the effective number of electrons scattered by Pt. The thickness of the headgroup is fixed to 4 Å. All other parameters are allowed the float within reasonable limits.

**Table S1.** X-ray fitting parameters for the data sets shown in Figure 2a.[a]

| Bulk Concentration and x-ray Energy | $\sigma_{subphase}$ (Å) | $\rho_{head}$ (e⁻/Å³) | $\rho_{head}(im)$ (e⁻/Å³) | $L_{head}$ (Å) | $\sigma_{head}$ (Å) | $\rho_{tail}$ (e⁻/Å³) | $L_{tail}$ (Å) | $\sigma_{tail}$ (Å) |
|---|---|---|---|---|---|---|---|---|
| 5μm $E_o$-250eV | 2.58 | 0.55 | 0 | 4* | 3.78 | 0.33 | 15.68 | 2.47 |
| 5μm $E_o$ | 2.49 | 0.54 | 0 | | | | | |
| 50μm $E_o$-250eV | 2.28 | 0.61 | 5.68E-04 | 4* | 3.2 | 0.33 | 16.51 | 2.69 |
| 50μm $E_o$ | 2.22 | 0.60 | 5.68E-04 | | | | | |
| 1 mM $E_o$-250eV | 3.11 | 0.72 | 3.90E-03 | 4* | 2.67 | 0.33 | 17.8 | 2.63 |
| 1 mM $E_o$ | 3.22 | 0.70 | 4.97E-03 | | | | | |
| 20 mM $E_o$-250eV | 3.68 | 0.79 | 1.49E-01 | 4* | 2.74 | 0.33 | 18.51 | 2.66 |
| 20 mM $E_o$ | 3.68 | 0.74 | 1.76E-01 | | | | | |

[a] The EDPs plotted in Figure 2b are based on these parameters. $\sigma$, $\rho$, and L represent the interfacial roughness, electron density, and thickness for each layer. The imaginary electron density ($\rho_{head}(im)$) is used for the absorption and only becomes non-zero at high Pt concentrations in the Stern layer. The absorption for other parts of the system is negligible. The subphase electron density is 0.354 e⁻/Å³ for all samples. *The thickness of the headgroups are fixed.

The difference between the on-edge and the off-edge measurements are caused by the number of effective electrons in Pt ions (Figure S1).[3] Therefore the difference can be used to determine the elemental EDP for Pt ions (Figure 2b inset of the main text). We can calculate the area under these curves to determine the area per Pt, considering that ~16 e⁻ corresponds to 1 Pt ion. (Figure 4, main text).

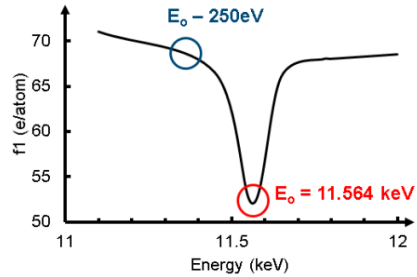

**Figure S1.** Effective number of electrons of Pt ions around the $L_3$ absorption edge. The blue and the red circles show the energies at which the a-XR measurements were done.

**Atomistic MD Simulations**

Classical MD simulations were performed using the GROMACS package (version 4.5.5). [4] The CHARMM 36 force field [5] was employed, which has been implemented under the GROMACS package. [6] The force field parameters of the $PtCl_6^{2-}$ have been reported by Lienke et al. in 2001, [7] which were developed in the framework of the CHARMM force field. Note that the same groups reported a new version of their force field in 2011. [8] Our simulations showed that these two sets of force fields provided quantitatively similar results regarding the surface activity of the two kinds of metalates investigated.

The CHARMM TIP3P water model was employed as in Ref. [7]. The water structure was constrained using the SETTLE algorithm.[9] The force field parameters of $H_3O^+$ reported by Sagnella and Voth [10] were employed. Given the fact that no van der Waals force field parameters for $Li^+$ ions were reported for the CHARMM force field, the corresponding parameters from the AMBER force field, [11] which uses the same combination rule (Lorentz−Berthelot rule[12-13]) as the CHARMM force field for the non-bonded Lennard-Jones 12-6 and Coulomb interactions, were employed instead. All the other parameters were from the original CHARMM 36 force field.

The initial structure were built using the package Packmol.[14] The lengths of the simulation box were 4.5×4.5×30 $nm^3$ in X×Y×Z dimensions. After the equilibration (see below for the details), a fixed lateral area 4.382×4.382 $nm^2$ was selected to meet the experimental area per lipid of around 0.48 $nm^2$ per DPTAP molecule. All the molecules, except DPTAP, were initially randomly distributed inside the water region of roughly 4.5×4.5×8 $nm^3$. The DPTAP molecules were located at the upper and the lower boundary of the water region with the hydrophilic headgroups extending inwards. The large vacuum region was included to mimic the experimental water/air biphasic condition (Figure S2).

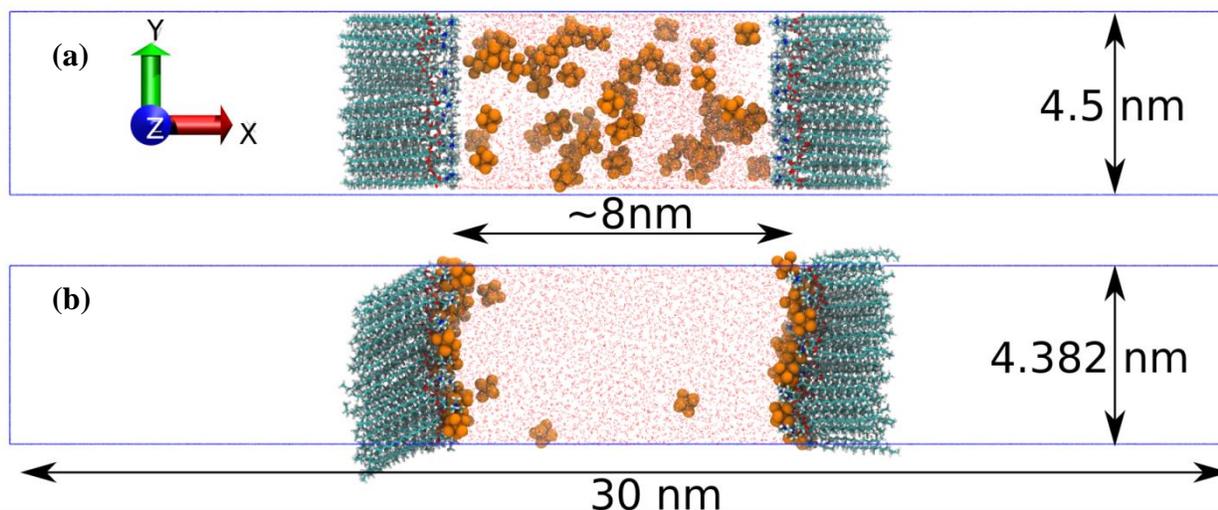

**Figure S2**. Snapshot of the (a) initial and (b) final structures of the 0.5 M $PtCl_6^{2-}$ aqueous solution in the water/air system. The $PtCl_6^{2-}$ ions are highlighted in orange. The lateral area of 4.5×4.5 $nm^2$ was initially employed for the convenience of the system preparation using Packmol, which was
S3

changed to be 4.382×4.382 nm$^2$ after the equilibration to meet the desired area per DPTAP of around 0.48 nm$^2$ (Figure 5b). Blue sold lines denote the simulation box boundary.

Due to the finite size effect in the MD simulations, it can be reasonably predicted that because of the adsorption at the water/DPTAP interface, the concentration of metalate in the central water region is lower than the total concentration. That is to say, the metalates will be distributed partially in the central water region, contributing to the "effective concentration" in the bulk water regime, and partially at the water/DPTAP interface regime. In this regard, we first simulated a series of aqueous solutions by varing the total concentrations of [PtCl$_6^{2-}$] = 0.1, 0.2, 0.4, 0.5, 0.6, 0.7, 0.8 M. See the following section for the simulation equilibration process. It was found that when [PtCl$_6^{2-}$] = 0.5 M, some metalate complexes stayed dissolved in the bulk water region dynamically, and the positive charges from the DPTAP headgroup were compensated (or slightly overcompensated) by the negative charges from the metalates. At concentrations lower than 0.5 M, all the metalates were distributed close to the water/DPTAP interface. The difference in the metalate adsorption at low concentration between the simulations and the experiments possibly indicates an underestimated hydration free energy in the simulations, which was not taken into account in the force field development.[7] In comparison, at concentrations higher than 0.5 M, a larger amount of metalates were distributed in the central water region. Therefore, in what follows, we will focus on the systems with the total concentration of 0.5 M metalates, which is the best description of the Stern layer in the 20 mM solution in the real experiments. Orders of magnitude difference in interfacial and bulk concentrations is expected in these systems. The limitations on the simulation box size prevents us from exactly matching both conditions, and therefore we focus on the interface.

**Table S2.** Numbers of the Components in the Simulation of 0.5M PtCl$_6^{2-}$

|  | [PtCl$_6$]$^{2-}$•K$_2^+$ [a] | Li$^+$•Cl$^-$ [b] | [H$_3$O]$^+$•Cl$^-$ [c] | DPTAP$^+$•Cl$^-$ [d] | H$_2$O |
|---|---|---|---|---|---|
| 0.5M PtCl$_6^{2-}$ | 50 | 50 | 1 | 80 | 5200 |

a. K$^+$ are counterions of metalates
b. [Li$^+$•Cl$^-$] = 0.5 M.
c. [H$_3$O$^+$•Cl$^-$] = 0.01 M (pH = 2)
d. Lateral area per DPTAP is 0.48 nm$^2$.

As aforementioned, all the molecules were initially randomly located in the water regime, with the DPTAP capped at the upper and lower boundaries in the Z-dimension (Figure S2a). The energy minimization of the initial structure was performed using the steepest descent algorithm. Each of the systems was subsequently equilibrated using semi-isotropic pressure coupling (P$_{XY}$ = P$_Z$ = 1 bar). The other simulation parameters were the same as those employed in the production simulations below. The equilibration simulation ran for a duration of 10 ns.

The lateral area of 4.382×4.382 nm$^2$ was then applied to reach the desired area per DPTAP of 0.48 nm$^2$. The following production simulation ran for 220 ns, with the simulation frames from the last 200 ns saved using a saving frequency of 10 ps per frame for the subsequent data analysis. In the production simulations, the NTV ensemble (constant number of particles, temperature, and volume) was used. The reference temperature was 298 K, with waters and the other molecules separately coupled using the velocity rescaling algorithm (time constant 0.1 ps). Three-dimensional periodic boundary conditions were employed. Neighbor searching was done up to a cutoff distance of 1.2 nm. The short-range Coulomb interactions were calculated up to this cutoff distance with the long-range Coulomb interactions calculated using the smooth Particle Mesh Ewald (PME) method with a grid real spacing of 0.12 nm and cubic interpolation.[15-16] The Lennard-Jones 12-6 potential was employed for the van der Waals interactions, which was calculated up to the cutoff distance of 1.2 nm, with the long-range dispersion correction for the energy and pressure applied. A simulation integration time step of 2 fs was employed with all the hydrogen-involved covalent bond lengths constrained using the LINCS algorithm.[17]

**Figure S3.** Electron density of the different components in the system with 0.5 M $PtCl_6^{2-}$. The

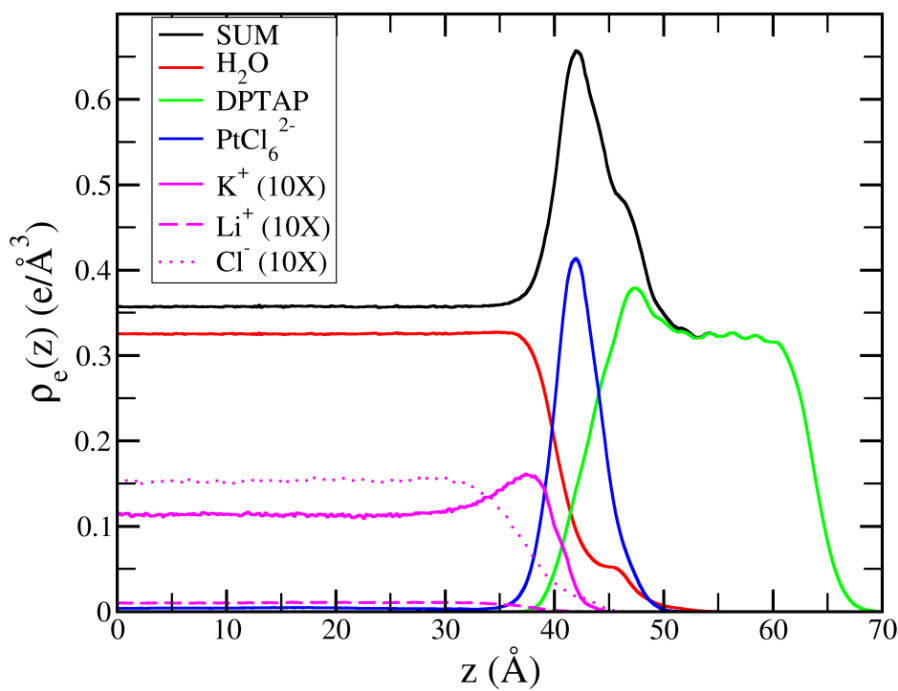

reference (z = 0) is based on the geometric center of all water molecules in the simulation box.

**SI References**